\documentclass[twocolumn,pra,showpacs,preprintnumbers,superscriptaddress,amsmath,amssymb,10pt,aps]{revtex4}

\usepackage{mathptmx}
\usepackage{subfigure}
\usepackage{psfrag,graphicx}
\usepackage{dcolumn}
\usepackage{amsmath,amssymb}
\usepackage{bm}
\usepackage{color}
\usepackage{latexsym}
\usepackage{epstopdf}
\usepackage{color}
\usepackage[english]{babel}
\usepackage{latexsym}
\usepackage{psfrag,graphicx}
\usepackage{subfigure}
\usepackage{amsmath}
\usepackage{amssymb}
\usepackage{amsfonts}
\usepackage{bm}
\usepackage{natbib}
\usepackage{epstopdf}
\usepackage{appendix}

\definecolor{mygrey}{gray}{0.35}
\definecolor{myblue}{rgb}{0.2,0.2,0.8}
\definecolor{myzard}{cmyk}{0,0,0.05,0}
\definecolor{mywhite}{rgb}{1,1,1}
\definecolor{mywhite}{rgb}{1,1,1}
\definecolor{myred}{rgb}{1,0.,0.3}

\usepackage[colorlinks=true,citecolor=myblue,linkcolor=myred]{hyperref}

\def\ba{\begin{align}}
\def\enda{\end{align}}
\def\bi{\begin{itemize}}
\def\ei{\end{itemize}}

\def\be{\begin{equation}}
\def\ee{\end{equation}}
\def\bea{\begin{eqnarray}}
\def\eea{\end{eqnarray}}
\def\bse{\begin{subequations}}
\def\ese{\end{subequations}}

\begin{document}
\title{Quantum metrology with critical driven-dissipative collective spin system}
\author{Venelin P. Pavlov}
\affiliation{Department of Physics, St. Kliment Ohridski University of Sofia, James Bourchier 5 blvd, 1164 Sofia, Bulgaria}
\author{Diego Porras}
\affiliation{Institute of Fundamental Physics IFF-CSIC, Calle Serrano 113b, 28006 Madrid, Spain}
\author{Peter A. Ivanov}
\affiliation{Department of Physics, St. Kliment Ohridski University of Sofia, James Bourchier 5 blvd, 1164 Sofia, Bulgaria}

\begin{abstract}
We propose a critical dissipaive quantum metrology schemes for single parameter estimation which are based on a quantum probe consisting of coherently driven ensemble of $N$ spin-1/2 particles under the effect of squeezed, collective spin decay. The collective spin system exhibits a dissipative phase transition between thermal and ferromagnetic phases, which is characterized with nonanalytical behavior of the spin observables. We show that thanks to the dissipative phase transition the sensitivity of the parameter estimation can be significantly enhanced. Furthermore, we show that our steady state is an entangled spin squeezed state which allow to perform parameter estimation with sub shot-noise limited measurement uncertainty.

\end{abstract}

%\pacs{
%03.67.Ac, %Quantum computation architectures and implementations
%03.67.Bg,
%03.67.Lx,
%42.50.Dv %Coherent control of atomic interactions with photons
%}
\maketitle

%%%%%%%%%%%%%%%%%%%%%%%%%%%%%%%%%%%%%%%%%%%%%%%%%%%%%%%%%%%%%%%%%%%%%%%%%%%
%%%%%%%%%%%%%%%%%%%%%%%%%%%%%%%%%%%%%%%%%%%%%%%%%%%%%%%%%%%%%%%%%%%%%%%%%%%
%%%%%%%%%%%%%%%%%%%%%%%%%%%%%%%%%%%%%%%%%%%%%%%%%%%%%%%%%%%%%%%%%%%%%%%%%%%
%========================================================================
%========================================================================
\section{Introduction}
Quantum metrology and quantum sensors are some of the most promising applications of quantum technologies \cite{Degen2017,Pezze2018,Giovannetti2011,Giovannetti2006}. Various areas of modern physics rely on the measurement of weak signals using quantum system as a probe. High-precision quantum sensors are of a key importance for testing gravity and its quantization and find wide applications in biology and medicine. One way to achieve greatly enhanced parameter estimation is to explore quantum critical systems as probes. Indeed, the emergence of the quantum phase transition can be used to enhanced the sensitivity of the quantum measurement by tuning the systems close to the critical point \cite{Zanardi2008,Liu2021,Ivanov2013,Salvatori2014,Felicetti2020,Chu2021,Rams,Pezze,Ying2022,Gietka2021,Ding,Zhou2023}. Usually, the critical quantum metrology protocols are based on either on an adiabatic evolution along the ground state energy or quantum quench where in both approaches the sensitivity is enhanced in the vicinity of the critical point.
%However, by approaching the critical point the energy gap is %closing which spoils the enhancement of the sensitivity since %it requires growth of the protocol duration.

A different class of critical non-equilibrium phenomena emerge in an open quantum systems where the competition between coherent driving and dissipation brings the system into a stationary state \cite{Kessler2012,Minganti2018}. The steady state is defined as an eigenmatrix of the Liouvillian superoperator with largest eigenvalue equal to zero. The dissipative phase transitions are associated with the existence of different steady states in the thermodynamic limit and the transition between them is characterized with a non-analytical behaviour of the steady state observable under the change of the control parameter. The dissipative phase transition is also characterized with vanishing of the Liouvillian gap, in which the second largest eigenvalue of the Liouvillian superoperator, often called asymptotic decay rate tends to zero at the critical point. Moreover, the experimental observation of dissipative phase transition has been realized in a various quantum-optical platforms including for example one-dimensional circuit QED lattice \cite{Fitzpatrick2017}, semiconductor microcavity \cite{Fink2017,Li2022}, and ultracold bosonic quantum gas \cite{Benary2022}. Such a critical driven-dissipative interaction may drives the system into entangled many-body stationary state and thus it can be used as a resource for high-precision quantum metrology \cite{Lorenzo2017,Lorenzo2018,Ivanov2020,Ivanov2020_1,Ilias2022,Macieszczak2016}.

In this work we study the metrological capability for measurement a small parameter using coherently driven ensemble of $N$ spin-1/2 particles with squeezed, collective spin decay. The model displays dissipative phase transition between ferromagnetic phase, which is characterizes with well defined spin magnetization, and respectively, a thermal phase which is characterized with zero magnetization \cite{Munoz}. Our quantum metrological scheme consists of two different approaches.
In the first approach the collective spin system is prepared in an arbitrary initial state which evolves toward an unique steady state. Then, a single shot measurement is performed which allows to determine one of the system parameters, namely, the amplitude $\Omega$ of the driving field or the squeezing angle $\theta$. The essence of this approach is an observation that in the vicinity of the critical point the steady state density matrix becomes sensitive to an infinitesimal small variation of the control parameter. We quantify the parameter sensitivity in terms of quantum Fisher information (QFI). We show that due to the onset of dissipative phase transition the estimation precision is greatly enhanced close to the critical point. Furthermore, we perform a finite size scaling analysis and show that at the critical line the measurement uncertainty can reach sub shot-noise limit of sensitivity where the QFI scales as $\sim N^{4/3}$. An important advantage of using critical dissipative quantum system as a probe is the lack of requirement for initial entangled state preparation as well as the needs to adiabatically follow a particular path in parameter space to bring the system close to the critical point. In the second approach the system is prepared initially in a steady state of the driven dissipative collective spin system and subsequently a parameter dependent unitary perturbation is applied. We study the criteria for particle entanglement and show that steady state is an entangled spin squeezed state. We show that the QFI obeys the scaling $\sim N^{4/3}$ as in the first approach. Thus, the two approaches can be used for high-precision quantum metrology with sensitivity beyond the standard shot noise limit.

The paper is organized as follows: In Sec. \ref{GF} we provide the theoretical framework for dissipative quantum metrology. Section \ref{system} presents the critical quantum probe consisting of a coherently driven ensemble of $N$ spin-1/2 particles with squeezed, collective spin decay. In Sec. \ref{quantum_metrology} we discuss our two quantum metrology protocols using critical driven dissipative quantum probe. We show that close to the dissipative phase transition one can achieve significant enhancement of the parameter sensitivity which overcomes the shot-noise limit.   Finally, the conclusions are presented in Sec. \ref{C}.

\section{Theoretical framework for parameter quantum estimation}\label{GF}
\subsection{General framework}
In order to estimate unknown parameter $\lambda$ one has to measure some observable $\hat{A}$, called an estimator. The error propagation formula which quantifies the estimation precision is given by the inverse of signal-to-noise ratio:
\begin{equation}
\Delta(\hat{A},\lambda)=\frac{\sqrt{\langle \hat{A}^{2}\rangle-\langle \hat{A}\rangle^{2}}}{\left|\frac{\partial\langle \hat{A}\rangle}{\partial\lambda}\right|},
\end{equation}
where $\langle \hat{A}\rangle={\rm Tr}(\hat{A}\hat{\rho}(\lambda))$ and $\hat{\rho}(\lambda)$ is the density matrix. Then the statistical uncertainty of the single parameter estimation is
\begin{equation}
\delta\lambda\geq \Delta(\hat{A},\lambda).\label{ccrb}
\end{equation}
For a quantum state with $N$ uncorrelated particles the error propagation formula scales as $\Delta(\hat{A},\lambda)\sim N^{-1/2}$ which is known as a \emph{shot-noise limit}. Quantum correlated states however may yield a favorable scaling in the parameter estimation. Indeed, using entangled states as a resource for quantum metrology can lead to a \emph{sub shot-noise sensitivity} where the error propagation formula obeys the scaling law $\Delta(\hat{A},\lambda)\sim N^{-a}$ with $a>1/2$.

The optimal strategy to measure $\lambda$ is usually associated with a privileged observable which minimize the statistical uncertainty and thus allows to estimate the unknown parameter with ultimate precision determined by the quantum Fisher information $\mathcal{F}_{\rm Q}(\lambda)$. The ultimate precision in the parameter estimation is quantified by the quantum Cramer-Rao (QCR) bound \cite{Degen2017,Pezze2018,Paris2009}
\begin{equation}
\delta\lambda\geq \mathcal{F}_{\rm Q}(\lambda)^{-1/2},\label{qcr}
\end{equation}
where we have $\Delta(\hat{A},\lambda)\geq \mathcal{F}_{\rm Q}(\lambda)^{-1/2}$. In order to find the QFI, we define the symmetric logarithmic derivative (SLD) operator $\hat{L}$ which obeys the operator equation $2\partial_{\lambda}\hat{\rho}(\lambda)=\hat{L}\hat{\rho}(\lambda)+\hat{\rho}(\lambda)\hat{L}$ such that $\mathcal{F}_{\rm Q}(\lambda)={\rm Tr}(\hat{\rho} \hat{L}^{2})$. Finally, the quantum Cramer-Rao bound can be saturated by the measurement projecting on the eigenvectors of SLD operator.
\subsection{Steady state quantum metrology}

In the following we consider a driven-dissipative quantum system whose dynamics obeys the master Lindblad equation \cite{Breuer}
\begin{equation}
\partial_{t}\hat{\rho}(t,\lambda)=\mathcal{L}\hat{\rho}(t,\lambda),
\end{equation}
where $\mathcal{L}$ is the Liouvillian superoperator which is trace preserving and generates a completely positive map $e^{\mathcal{L} t}$ describing the time evolution of the system. In general, the interplay between the coherent dynamics and the dissipative processes give rise to a steady state solution of the master Lindblad equation determined by $\hat{\rho}_{\rm ss}(\lambda)=\lim_{t \to \infty}e^{\mathcal{L} t} \hat{\rho}_{\rm in}$ and $\mathcal{L}\hat{\rho}_{\rm ss}(\lambda)=0$. The latter equality implies that the steady state is an eigenmatrix of $\mathcal{L}$ with zero eigenvalue. A steady state observable may shows discontinuity in the thermondynamic limit $N\rightarrow\infty$ which is associated with the onset of a dissipative phase transition \cite{Kessler2012,Minganti2018}. In the following we explore the onset of dissipative phase transition in a driven-dissipative collective spin system for enhanced parameter estimation. Our quantum metrology schemes consist of two approaches, namely:

(i) In the first quantum metrology scheme which we refer as a \emph{critically-enhanced steady state parameter estimation}, the system is prepared initially in an arbitrary state which evolves under the action of the master Lindblad equation towards the steady state $\hat{\rho}_{\rm ss}(\lambda)$ which possesses a spectral decomposition $\hat{\rho}_{\rm ss}(\lambda)=\sum_{n}p_{n}|\psi_{n}\rangle\langle\psi_{n}|$, where $p_{n}$ and $|\psi_{n}\rangle$ are the eigenvalues and eigenvectors of the steady state density matrix. Subsequently, the observable $\hat{A}$ is measured which allows to estimate the unknown parameter $\lambda$ with statistical unsertainty given by Eq. (\ref{ccrb}). The ultimate precision is bounded by the QCR inequality (\ref{qcr}) where the QFI can be expressed as
\begin{equation}
\mathcal{F}_{\rm Q}(\lambda)=2\sum_{n,m}\frac{\langle\psi_{n}|\partial_{\lambda}\hat{\rho}_{\rm ss}|\psi_{m}\rangle\langle\psi_{m}|\partial_{\lambda}\hat{\rho}_{\rm ss}|\psi_{n}\rangle}{p_{n}+p_{m}},\label{vqfi}
\end{equation}
with $p_{n}+p_{m}>0$ and for simplicity we write $\hat{\rho}_{\rm ss}(\lambda)=\hat\rho_{\rm ss}$. We will show that the sensitivity of the parameter estimation can be significantly enhanced close to a dissipative phase transition. Furthermore, at the dissipative critical point we can reach a sub shot-noise sensitivity of the parameter estimation where the QFI obeys the scaling $\mathcal{F}_{\rm Q}(\lambda)\sim N^{4/3}$.

(ii) In the second approach which we refer as a \emph{perturbed steady state parameter estimation} the system is prepared in the steady state with density matrix $\hat{\rho}_{\rm ss}$. Then the system evolves unitary into a final state with density matrix $\hat{\rho}_{\rm f}(\lambda)=e^{-i\lambda \hat{G}}\hat{\rho}_{\rm ss} e^{i\lambda \hat{G}}$ where $\hat{G}$ is a Hermitian operator. The QFI is given by
\begin{equation}
\mathcal{F}_{\rm Q}(\lambda)=2\sum_{n,m}\frac{(p_{n}-p_{m})^{2}}{p_{n}+p_{m}}|\langle\psi_{n}|\hat{G}|\psi_{m}\rangle|^{2}.\label{qfi2}
\end{equation}
The driven-dissipative dynamics can be used as a resource for entanglement \cite{Torre2013,Tudela2013,Groszkowski2022}. In order to quantify the condition for particle entanglement in the steady state density matrix we rewrite the QCR bound as \cite{Pezze2009}
\begin{equation}
\delta\lambda\geq\frac{\chi}{\sqrt{N}},
\end{equation}
where $\chi^2=N/\mathcal{F}_{\rm Q}(\lambda)$. Thus a necessary and sufficient condition to achieve a sub shot-noise sensitivity of the parameter estimation is $\chi^2<1$. We will show that our steady state is entangled spin squeezed state. Moreover, at the critical point the QFI possesses the scaling $\mathcal{F}_{\rm Q}(\lambda)\sim N^{4/3}$ as in the first approach. Therefore, the two quantum metrology schemes can be used for high-precision parameter estimation beyond the standard quantum limit.

In the following we discuss the two quantum metrology approaches discussed above using driven-dissipative collective spin system which exhibits dissipative phase transition between thermal to ferromagnetic phases.

\section{Driven-dissipative spin system}\label{system}
We consider a coherently-drive ensemble of $N$ spin-1/2 particles with squeezed, collective spin decay. The master Lindblad equation of the system reads
\begin{equation}
\partial_{t}\hat{\rho}=-i\Omega[\hat{S}_{x},\hat{\rho}]+\frac{\Gamma}{N}\mathcal{D}_{\hat{S}_{\theta}}[\hat{\rho}],\label{master}
\end{equation}
where $\Omega$ is the driving frequency and $\Gamma$ is the quantum-jump rate. The Lindblad superoperator is $\mathcal{D}_{\hat{S}_{\theta}}[\hat{\rho}]=2\hat{S}_{\theta}\hat{\rho}\hat{S}_{\theta}^{\dag}-\{\hat{S}_{\theta}^{\dag}\hat{S}_{\theta},\hat{\rho}\}$, where the quantum jump operator $\hat{S}_{\theta}=\cos(\theta)\hat{S}_{-}+\sin(\theta)\hat{S}_{+}$ includes both lowering and raising collective spin operators $\hat{S}_{\pm}$ parametrized by a squeezing angle $\theta$. For angle $\theta=0$ the model corresponds to the standard case of collective resonance fluorescence \cite{Puri1979,Lawande1981}. Furthermore, the jump operator $\hat{S}_{\theta}$ possesses a dark state which is an entangled spin-squeezed state \cite{Torre2013}.
\begin{figure}
\includegraphics[width=0.45\textwidth]{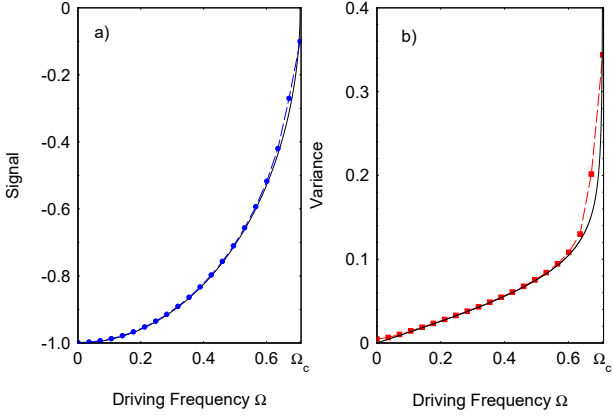}
\caption{(a) The expectation value $\langle \hat{s}_{z}\rangle$ at the steady state as a function of the driving frequency $\Omega$. We compare the mean value derived from the master Lindblad equation (\ref{master}) (dashed blue circles) with the analytical expression (solid line). The parameters are set to $\Gamma=1$, $\theta=\pi/8$, and $N=300$. (b) Exact result for the variance of the signal $\sqrt{\Delta \hat{s}_{z}^{2}}$ (dashed red squares) compared with the analytical result (solid line).}
\label{fig1}
\end{figure}

The dynamics in Eq. (\ref{master}) can be realized in a driven spin ensemble coupled to a single cavity mode. In the limit of fast cavity decay the bosonic field tends to a stationary vacuum state and thus its adiabatic elimination can gives rise to an effective, collective spin dissipation \cite{Torre2013,Tudela2013}.

\subsection{Mean field solution}

The competition between the driven coherent dynamics and the processes of dissipation drives the system toward a nonequilibrium steady state. In order to describe the spin mean orientation, spin fluctuations, and the metrological properties of the system in the steady state regime we express the collective spin operators in terms of a single bosonic mode with annihilation operator $\hat{b}$, using Holsten-Primakoff (HP) transformation where $\hat{S}_{+}=\hat{b}^{\dag}\sqrt{2S-\hat{b}^{\dag}\hat{b}}$ and $\hat{S}_{z}=\hat{b}^{\dag}\hat{b}-S$ with $S=N/2$ being the collective spin length. To account the spin mean polarization we displace the bosonic operators via $\hat{b}\rightarrow \hat{b}+\sqrt{S}\beta$. We also normalize the operators according to $\hat{s}_{-}=\hat{S}_{-}/S$ and $\hat{s}_{z}=\hat{S}_{z}/S$. Then using the HP representation we can expand the collective spin operators in a power series of $\epsilon=1/\sqrt{S}$, namely $\hat{s}_{-}=\sum_{l=0}^{\infty}\epsilon^{l}\hat{s}_{-}^{(l)}$ where up to first order in $\epsilon$ we have
\begin{equation}
\hat{s}_{-}^{(0)}=\sqrt{k}\beta,\quad \hat{s}^{(1)}_{-}=A \hat{b}+B \hat{b}^{\dag},
\end{equation}
with $A=(2k-|\beta|^{2})/2\sqrt{k}$, $B=-\beta^{2}/2\sqrt{k}$, and $k=2-|\beta|^{2}$. Similarly, for $\hat{s}_{z}$ operator we have $\hat{s}_{z}=\sum_{l=0}^{2}\epsilon^{l} s^{(l)}_{z}$ with
\begin{equation}
\hat{s}^{(0)}_{z}=|\beta|^{2}-1,\quad \hat{s}^{(1)}_{z}=\beta \hat{b}^{\dag}+\beta^{*}\hat{b},\quad \hat{s}^{(2)}_{z}=\hat{b}^{\dag}\hat{b}.
\end{equation}
We also expand the density matrix in a power series, $\hat{\rho}=\sum_{l=0}^{\infty}\epsilon^{l}\hat{\rho}_{l}$ with the requirement that ${\rm Tr}\hat{\rho}_{0}=1$ and ${\rm Tr}\hat{\rho}_{l}=0$ for $l\neq 0$. Then grouping the terms of power of $\epsilon$ in (\ref{master}) we obtain an equation for the displacement parameter $\beta$
\begin{equation}
[\hat{s}^{(1)}_{+}(-i\Omega-\hat{s}^{(0)}_{-}\Omega_{\rm c})+\hat{s}^{(1)}_{-}(-i\Omega+\hat{s}^{(0)}_{+}\Omega_{\rm c}),\hat{\rho}_{0}]=0,
\end{equation}
\begin{figure}
\includegraphics[width=0.45\textwidth]{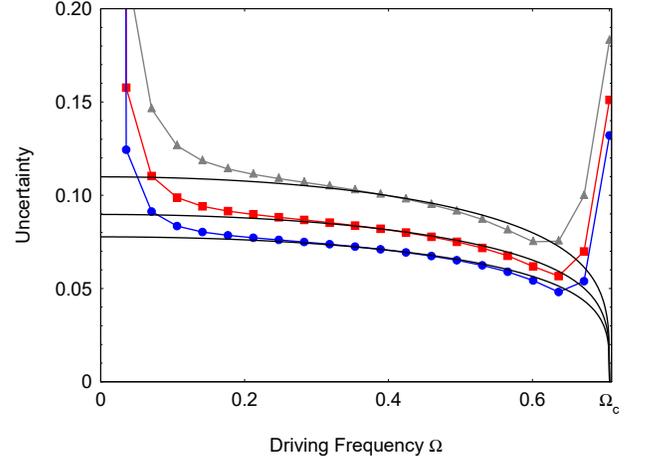}
\caption{Statistical uncertainty $\delta\Omega$ bounded by the error propagation formula versus $\Omega$. We compare the exact result derived from Eq. (\ref{master}) for $N=100$ (grey triangles line), $N=150$ (red squares line), $N=200$ (blue circles line) with the analytical result (\ref{unc}) (solid lines). The other parameters are $\Gamma=1$ and $\theta=\pi/8$.}
\label{fig2}
\end{figure}which gives $\beta=-i\sqrt{1-M}$ with $M=\sqrt{1-(\Omega/\Omega_{\rm c})^{2}}$ and $\Omega_{\rm c}=\Gamma\cos(2\theta)$ being the critical coupling. Finally, the master equation for the lowest order density matrix $\hat{\rho}_{0}$ is given by
\begin{eqnarray}
\partial_{t}\hat{\rho}_{0}&=&\frac{\gamma_{-}}{2}\mathcal{D}_{\hat{b}}[\hat{\rho}_{0}]+
\frac{\gamma_{+}}{2}\mathcal{D}_{\hat{b}^{\dag}}[\hat{\rho}_{0}]+\frac{\eta}{2}(2\hat{b}\hat{\rho}_{0}\hat{b}-\{\hat{b}^{2},\hat{\rho}_{0}\})\notag\\
&&+\frac{\eta}{2}(2\hat{b}^{\dag}\hat{\rho}_{0}\hat{b}^{\dag}-\{b^{\dag 2},\hat{\rho}_{0}\}),\label{rho0}
\end{eqnarray}
where $\gamma_{-}=\Gamma_{-}A^{2}+\Gamma_{+}B^{2}+2\xi AB$, $\gamma_{+}=\Gamma_{+}A^{2}+\Gamma_{-}B^{2}+2\xi AB$, $\eta=AB(\Gamma_{-}+\Gamma_{+})+\xi(A^{2}+B^{2})$ and $\Gamma_{-}=\Gamma\cos^{2}(\theta)$, $\Gamma_{+}=\Gamma\sin^{2}(\theta)$, $\xi=\Gamma\sin(2\theta)/2$.

In the stationary limit $t\rightarrow\infty$ the steady state values of the observables are obtained by setting the time derivative of Eq. (\ref{rho0}) to zero. Then, the mean value of the observable $\hat{A}$ at the lowest order of $\epsilon$ is defined by $\langle \hat{A}\rangle={\rm Tr}(\hat{A}\hat{\rho}_{\rm ss})$, where $\hat{\rho}_{\rm ss}$ is the steady state density matrix.
\subsection{Gaussian state}
Because the dynamics for the lowest order density matrix $\hat{\rho}_{0}$ is quadratic in the bosonic operators, the steady state of the system is of single mode Gaussian form and the density matrix $\hat{\rho}_{\rm ss}$ can be reconstructed from the first and the second moments. In order to find the single mode Gaussian state of the system, we define quadrature operator $\hat{\textbf{q}}=\{\hat{x},\hat{p}\}^{\rm T}$ and mean displacement vector $\textbf{d}=\langle \hat{\textbf{q}}\rangle$, where $\hat{x}=(\hat{b}^{\dag}+\hat{b})$ and $\hat{p}=i(\hat{b}^{\dag}-\hat{b})$ are the position and momentum operators of the bosonic mode. Then, the covariance matrix is given by
\begin{equation}
\Sigma_{kl}(\hat{\rho}_{\rm ss},\hat{\textbf{q}})=\frac{1}{2}\langle \hat{q}_{k}\hat{q}_{l}+\hat{q}_{l}\hat{q}_{k}\rangle-d_{k}d_{l}.
\end{equation}
A general single-mode Gaussian state can be expressed as a squeezed displaced thermal state \cite{Weedbrook2012,Pinel2013},
\begin{equation}
\hat{\rho}_{\rm ss}=\hat{R}(\delta)\hat{D}(\alpha)\hat{S}(\zeta)\hat{\rho}_{\rm th}\hat{S}^{\dag}(\zeta)\hat{D}^{\dag}(\alpha)\hat{R}^{\dag}(\delta),
\end{equation}
where $\hat{R}(\delta)=e^{i\delta \hat{b}^{\dag}\hat{b}}$ is the rotation operator, $\hat{D}(\alpha)=e^{\alpha(\hat{b}^{\dag}-\hat{b})}$ is the displacement operator, and $\hat{S}(\zeta)=e^{\frac{r}{2}(\hat{b}^{2}e^{-2i\phi}-\hat{b}^{\dag 2}e^{2i\phi})}$ is the squeezing operator. Finally, the thermal state density matrix is $\hat{\rho}_{\rm th}=\sum_{n}p_{n}|n\rangle\langle n|$, where $p_{n}=N^{n}_{\rm th}/(1+N_{\rm th})^{n+1}$, with $N_{\rm th}$ being the mean number of thermal excitations.
\begin{figure}
\includegraphics[width=0.45\textwidth]{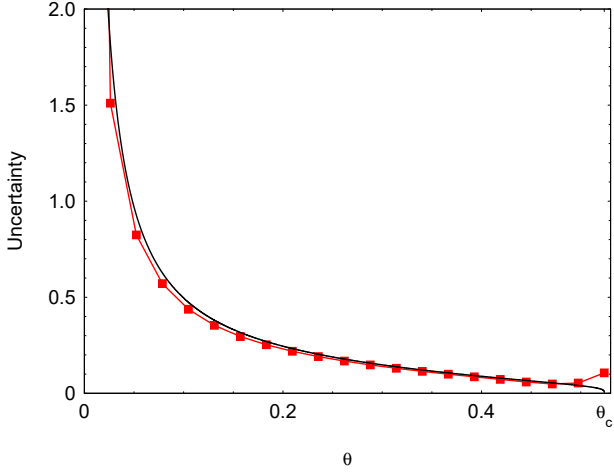}
\caption{Uncertainty of the phase estimation. We compare the exact result (red squares line) with the analytical expression (\ref{unc_phase}). The parameters are set to $\Omega=0.5$, $\Gamma=1$, and $N=100$.}
\label{fig4}
\end{figure}
We find that the covariance matrix elements in the steady state are given by
\begin{equation}
\Sigma_{11}=\frac{\gamma_{+}+\gamma_{-}-2\eta}{\gamma_{-}-\gamma_{+}},\quad \Sigma_{22}=\frac{\gamma_{+}+\gamma_{-}+2\eta}{\gamma_{-}-\gamma_{+}},\quad \Sigma_{12}=0.
\end{equation}
The purity of the steady state is $P=({\rm det}\Sigma(\hat{\rho}_{\rm ss},\hat{\textbf{q}}))^{-1/2}=(2N_{\rm th}+1)^{-1}$. Then it is straightforward to show that ${\rm det}\Sigma(\hat{\rho}_{\rm ss},\hat{\textbf{q}})=1$. Therefore, we have $N_{\rm th}=0$ and the steady state is a pure state. The displacement vector can be expressed as $\textbf{d}=2\alpha[\cos(\delta),\sin(\delta)]^{\rm T}$. Because in the steady state regime we have $\langle \hat{b}^{\dag}\rangle=\langle \hat{b}\rangle=0$, the displacement amplitude is $\alpha=0$. For the rest of parameters we have \cite{Weedbrook2012,Pinel2013}
\begin{eqnarray}
e^{-2r}\cos^{2}(\phi+\delta)+e^{2r}\sin^{2}(\phi+\delta)&=&\Sigma_{11},\notag\\
e^{2r}\cos^{2}(\phi+\delta)+e^{-2r}\sin^{2}(\phi+\delta)&=&\Sigma_{22},\notag\\
\sinh(2r)\sin(2\phi+2\delta)&=&\Sigma_{12}.
\end{eqnarray}
Therefore, we find that the squeezing parameter is
\begin{equation}
r=\frac{1}{2}\ln\left(\frac{1+M}{2\Omega_{\rm c}M}\left(\sqrt{\Gamma_{-}}+\sqrt{\Gamma_{+}}\right)^{2}\right).
\end{equation}
Finally, using that $\hat{R}(\delta)\hat{S}(\zeta)\hat{R}^{\dag}(\delta)=e^{\frac{r}{2}(\hat{b}^{2}e^{-2i(\phi+\delta)}-\hat{b}^{\dag 2}e^{2i(\phi+\delta)})}=\hat{S}(r)$ we find that the steady state is a squeezed state,
\begin{equation}
\hat{\rho}_{\rm ss}=\hat{S}(r)|0\rangle\langle 0|\hat{S}^{\dag}(r).
\end{equation}
We see that, in the thermodynamic limit the squeezing parameter diverges, in the vicinity of the critical line, where $M\rightarrow 0$. This leads to a non-analytical behaviour of the steady state and respectively to an onset of dissipative phase transition between ferromagnetic phase $\Omega<\Omega_{\rm c}$ which is characteriezed with well defined magnetization ($M\neq 0$) and respectively a thermal phase $\Omega>\Omega_{\rm c}$ which is characterized with zero magnetization, ($M=0$). Note that in the thermal phase the steady state is close to the infinite-temperature state $\hat{\rho}_{\rm ss}\propto\hat{\mathbb{I}}$ and thus this dissipative phase is not suitable for enhanced parameter estimation.

In the following we study the metrological properties of the driven-dissipative spin system focusing on the \emph{ferromagnetic phase}. We will show that close to the critical point we can achieve greatly enhanced parameter estimation.
\begin{figure}
\includegraphics[width=0.45\textwidth]{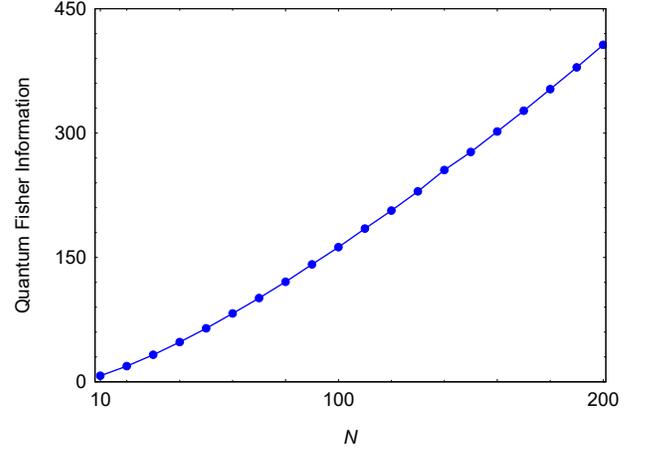}
\caption{Exact result of the QFI as a function of the number of spins $N$ at the critical coupling $\Omega=\Omega_{\rm c}$. The parameters are set to $\Gamma=1$, $\theta=\pi/8$. The fit of QFI gives $\mathcal{F}_{\rm Q}(\Omega_{\rm c})\sim a N^{b}$ with $a=0.34$ and $b=1.34$.}
\label{fig3}
\end{figure}

\section{Quantum Metrology Protocol}\label{quantum_metrology}

\subsection{Critically-enhanced steady state parameter estimation}
Let us now discuss the first approach for parameter estimation in which the information for the unknown parameter $\lambda$ is encoded in the steady state of the collective spin system. Then the parameter estimation is performed by measuring a suitable observable $\hat{A}$. Consider for example $\hat{A}=\hat{s}_{\mu}$ ($\mu=x,y,z$). For $\mu=x$ the corresponding collective spin observable is zero in the steady state, $\langle \hat{s}_{x}\rangle=0$, and thus it is not suitable for parameter estimation. The mean value of the other two observable are given by
\begin{eqnarray}
&&\langle \hat{s}_{z}\rangle\approx{\rm Tr}(\hat{s}_{z}^{(0)}\hat{\rho}_{\rm ss})+\epsilon^{2}{\rm Tr}(\hat{s}^{(2)}_{z}\hat{\rho}_{\rm ss})=-\sqrt{1-\frac{\Omega^{2}}{\Omega^{2}_{\rm c}}}+\mathcal{O}(\epsilon^{2}),\notag\\
&&\langle \hat{s}_{y}\rangle\approx{\rm Tr}(\hat{s}_{y}^{(0)}\hat{\rho}_{\rm ss})+\epsilon^{2}{\rm Tr}(\hat{s}^{(2)}_{y}\hat{\rho}_{\rm ss})=\frac{\Omega}{\Omega_{\rm c}}+\mathcal{O}(\epsilon^{2}).\label{signal}
\end{eqnarray}
We also find that the variance of the two observable in the steady state are
\begin{eqnarray}
&&\Delta \hat{s}^{2}_{z}\approx\epsilon^{2}{\rm Tr}(\hat{s}^{(1)2}_{z}\hat{\rho}_{\rm ss})=\epsilon^{2}\frac{1-M^{2}}{2\Omega_{\rm c}M}(\sqrt{\Gamma_{-}}+\sqrt{\Gamma_{+}})^{2}+\mathcal{O}(\epsilon^{4}),\notag\\
&&\Delta \hat{s}^{2}_{y}\approx\epsilon^{2}{\rm Tr}(\hat{s}^{(1)2}_{y}\hat{\rho}_{\rm ss})=\epsilon^{2}\frac{M}{2\Omega_{\rm c}}(\sqrt{\Gamma_{-}}+\sqrt{\Gamma_{+}})^{2}+\mathcal{O}(\epsilon^{4}).\label{variance}
\end{eqnarray}
In Fig. \ref{fig1} we show numerical result for finite system size for the signal $\langle \hat{s}_{z}\rangle$ and the variance $\sqrt{\Delta \hat{s}^{2}_{z}}$ compared with the analytical expressions Eqs. (\ref{signal}) and (\ref{variance}), where very good agreement is observed. Consider that the parameter we wish to estimate is the driving frequency $\lambda=\Omega$. Then, we find that the uncertainty of the parameter estimation for each of the observables is given by
\begin{equation}
\delta\Omega\geq\frac{1}{\sqrt{N}}(\Omega^{2}_{\rm c}-\Omega^{2})^{\frac{1}{4}}(\sqrt{\Gamma_{-}}+\sqrt{\Gamma_{+}}).\label{unc}
\end{equation}
Equation (\ref{unc}) indicates that by approaching the dissipative phase transition the parameter estimation is significantly improved. Indeed, in this limit, the quantum probe becomes sensitive to an infinitely small driving field. Figure (\ref{fig2}) shows comparison between the exact and analytical results for the statistical uncertainty (\ref{unc}). We see that by increasing $N$ the numerical result closely follow the analytical expression (\ref{unc}).

Alternatively, we can set $\lambda=\theta$. Then, the error propagation formula for each of the observables gives
\begin{equation}
\delta\theta\geq\frac{1}{2\sqrt{N}}(\Omega^{2}_{\rm c}-\Omega^{2})^{\frac{1}{4}}\frac{\sqrt{\Gamma_{-}}+\sqrt{\Gamma_{+}}}{\Omega\tan(2\theta)}\label{unc_phase},
\end{equation}
where we observe again enhancement of the phase estimation close to the optimal phase given by $\theta_{\rm c}=\frac{1}{2}\cos^{-1}(\Omega/\Gamma)$ as is shown in Fig. \ref{fig4}.

In general the error propagation formula is expected to scale away from the critical coupling as \cite{Rams}
\begin{equation}
\Delta(\hat{A},\Omega)\sim N^{-1/2}|\Omega_{\rm c}-\Omega|^{-\frac{d\nu}{2}+1},
\end{equation}
where $\nu$ is the critical exponent and $d$ is the spatial dimension. Similarly, away from the critical coupling the QFI is expected to scale as  $\mathcal{F}_{\rm Q}(\Omega)\sim N$ which leads to a shot noise sensitivity with respect to the number of spins. Using Eq. (\ref{unc}) we find that $d\nu=3/2$. Furthermore, the QFI is expected to scale at the critical point as
\begin{equation}
\mathcal{F}^{1/2}_{\rm Q}(\Omega_{\rm c})\sim N^{1/d\nu},
\end{equation}
which implies that $\mathcal{F}^{1/2}_{\rm Q}(\Omega_{\rm c})\sim N^{2/3}$ and hence one can overcome the shot-noise limit at the critical coupling. In Fig. \ref{fig3} we show the exact result for the QFI using Eq. (\ref{vqfi}). We see indeed that at the critical point $\Omega=\Omega_{\rm c}$ the QFI scales with the number of spins $N$ approximately as $\mathcal{F}_{\rm Q}(\Omega_{\rm c})\sim N^{4/3}$ such that we can achieve a sub shot noise sensitivity of the parameter estimation.
\begin{figure}
\includegraphics[width=0.45\textwidth]{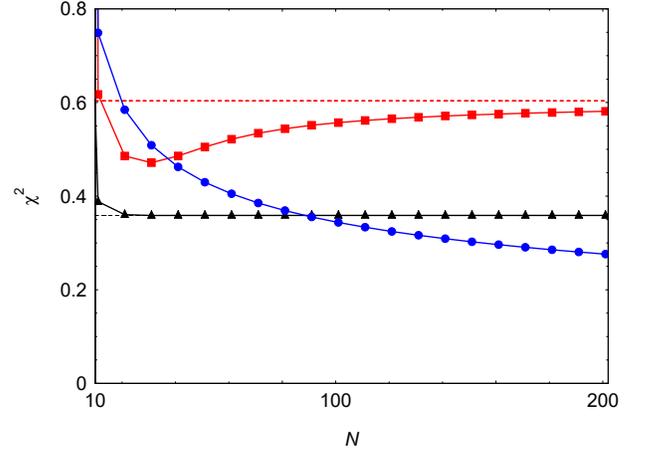}
\caption{Exact result for parameter $\chi^{2}$ versus $N$. We set $\Gamma=1$, $\theta=\pi/8$ and vary $\Omega$. The blue circle line corresponds to $\Omega=\Omega_{\rm c}$ and red squares line is for $\Omega=0.7\Omega_{\rm c}$ with $\hat{S}_{n}=\hat{S}_{z}$. The black triangle line is for $\Omega=0.5\Omega_{\rm c}$ and $\hat{S}_{n}=M \hat{S}_{y}+\sqrt{1-M^{2}}\hat{S}_{z}$. The dashed lines are the analytical solution in the thermodynamic limit with QFI given by Eq. (\ref{QFI3}). }
\label{fig5}
\end{figure}

\subsection{Perturbed steady state parameter estimation}
In the second quantum metrology approach the system is prepared in the steady state, which subsequently is perturbed by the unitary operator $\hat{U}_{\lambda}=e^{-i\lambda \hat{G}}$, where $\lambda$ is the parameter we wish to estimate. In the following we choose the generator to be $\hat{G}=\hat{S}_{n}=\hat{\vec{S}}.\vec{n}$, which is the projection of the collective spin operator along the direction $\vec{n}$. A reduced fluctuation along one of the spin directions provides enhancement of the parameter sensitivity over the shot-noise limit. The steady state $\hat{\rho}_{\rm ss}$ is a squeezed spin state as long as it displace a reduced fluctuation. A popular figure of merit which quantifies the degree of spin squeezing is \cite{Wineland,Ma2011}:
\begin{equation}
\xi^2=\frac{N(\Delta S_{\vec{n}_{\perp}})^2}{ |\langle S_{\vec{n}^{'}} \rangle|^{2} },
\end{equation}
where $\vec{n}^{'}$ is the collective spin mean direction and $\vec{n}_\perp$ is an arbitrary direction perpendicular to it. As long as there exist a direction $\vec{n}_{\perp}$ such that $\xi^2 < 1$ the quantum state with $N$ spins is a spin squeezed. Using Eq. (\ref{rho0}) the steady state collective spin mean direction in the ferromagnetic phase is given by $\vec{n}^{'}=\sqrt{1-M^2}\vec{y}-M\vec{z}$. Therefore, the optimal direction for which $\xi^2 < 1$ is the $\vec{x}$ axis.A different measure for particle entanglement is the parameter $\chi^2=N/\mathcal{F}_{\rm Q}(\lambda)$ where $\mathcal{F}_{\rm Q}(\lambda)$ is the QFI given by Eq. (\ref{qfi2}). A sufficient condition for multiparticle entanglement is $\chi^2<1$ which guaranties that the estimation of the phase shift $\lambda$ is with sensitivity $\delta\lambda$ beyond the shot noise. Since the steady state in the thermodynamic limit is a pure state we can evaluate the QFI. Indeed, we have
\begin{equation}
\mathcal{F}_{\rm Q}(\lambda)=4S^{2}\Delta \hat{s}^{2}_{\vec{n}}.\label{QFI3}
\end{equation}
For example, let us choose a direction $\vec{n}$ which is perpendicular to the collective
mean spin direction $\vec{n}^{'}$ which turns out to be optimal \cite{LMG_model}. Then we have $\hat{S}_{n}=M\hat{S}_{y}+\sqrt{1-M^2}\hat{S}_{z}$ and the QFI is given by
\begin{equation}
\mathcal{F}_{\rm Q}(\lambda)=\frac{N}{\Omega_{\rm c}M}(\sqrt{\Gamma_{-}}+\sqrt{\Gamma_{+}})^2.
\end{equation}
In Fig. (\ref{fig5}) we show the exact result for $\chi^2$ as a function of $N$ and for various $\Omega$. Away from the critical point ($\Omega<\Omega_{\rm c}$) the parameter is $\chi^{2}<1$ and respectively in the thermodynamic limit approaches
\begin{equation}
\chi^{2}=\frac{\Omega_{\rm c}M}{(\sqrt{\Gamma_{-}}+\sqrt{\Gamma_{+}})^{2}}.
\end{equation}
Therefore, preparing the system in the steady state, one can improve the sensitivity $\delta\lambda$ beyond the shot noise limit. Note that for $\Omega=0$ and $\theta=0$ we have $\chi=1$ which corresponds to the shot-noise limit. Crucially, approaching the critical point the spin magnetization tends to $M\rightarrow 0$ and respectively the QFI diverges, which signals the onset of dissipative phase transition. Hence, for a steady state close to the critical point the parameter estimation can be tuned to arbitrary high precision. Furthermore, the numerical result indicates that at $\Omega=\Omega_{\rm c}$ the parameter $\chi$ decreases with $N$ as $\chi^{2}\sim N^{-1/3}$ and thus $\mathcal{F}_{\rm Q}(\lambda)\sim N^{4/3}$. Therefore, the QFI obeys the same scaling as in the first approach.

\section{Conclusion}\label{C}
In summary, we have discussed quantum metrology schemes for single parameter estimation based on a dissipative phase transition of coherently driven collective spin system under the effect of squeezed, collective spin decay. The system exhibits dissipative phase transition between ferromagnetic phase which is characterized with well define collective spin magnetization, and respectively thermal phase which is characterized with zero magnetization. In the first approach an arbitrary initial state evolves toward an unique steady state which displays a nonanalytical behaviour in the thermodynamic limit at the critical coupling. Then, the parameter estimation is carried on by measuring one of the collective spin observables, which allow to determine one of the system parameters, namely the amplitude of the driving field or the squeezing angle. We have shown that thanks to the driven dissipative dynamics, one can achieve significant enhancement of parameter estimation close to the critical coupling, even for finite number of spins. Furthermore, we have shown that at the critical coupling, the QFI scales with the number of spins as $\mathcal{F}_{\rm Q}(\lambda)\sim N^{4/3}$, which leads to sub shot-noise sensitivity. In the second quantum metrology approach the system is prepared initially in the steady state of the collective spin system. Subsequently, the parameter dependent unitary perturbation is applied which drives the system into the final state. We have studied the criteria for particle entanglement and show that in the ferromagnetic phase $\chi^{2}<1$ which manifests that the steady state is an entangled spin squeezed state. We have quantified the estimation precision using QFI and show that obeys scaling $\mathcal{F}_{\rm Q}(\lambda)\sim N^{4/3}$ as in the fist approach. Hence, the two quantum metrology approaches based on a driven-dissipative quantum probe can be used for parameter estimation beyond the shot-noise limit.

%%%%%%%%%%%%%%%%%%%%%%%
\section*{Acknowledgments}
V. P. P. and P. A. I. acknowledge support by the Bulgarian national plan for recovery and resilience, contract BG-RRP-2.004-0008-C01 (SUMMIT: Sofia University Marking Momentum for Innovation and Technological Transfer), project number 3.1.4.


\begin{thebibliography}{99}

\bibitem{Degen2017} C. L. Degen, F. Reinhard, and P. Cappellaro, Rev. Mod. Phys. \textbf{89}, 035002 (2017).

\bibitem{Pezze2018} L. Pezz\'e, A. Smerzi, M. K. Oberthaler, R. Schmied, and P. Treutlein, Rev. Mod. Phys. \textbf{90}, 035005 (2018).

\bibitem{Giovannetti2011} V. Giovannetti, S. Lloyd, and L. Maccone, Nat. Photonics \textbf{5}, 222 (2011).

\bibitem{Giovannetti2006} V. Giovannetti, S. Lloyd, and L. Maccone, Phys. Rev. Lett. \textbf{96}, 010401 (2006).

\bibitem{Zanardi2008} P. Zanardi, M. G. A. Paris, and L. C. Venuti, Phys. Rev. A \textbf{78}, 042105 (2008).

\bibitem{Liu2021} R. Liu, Y. Chen, M. Jiang, X. Yang, Z. Wu, Y. Li, H. Yuan, X. Peng, and J. Du, npj Quantum Information \textbf{7}, 1 (2021).

\bibitem{Ivanov2013} P. A. Ivanov and D. Porras, Phys. Rev. A \textbf{88}, 023803 (2013).

\bibitem{Salvatori2014} G. Salvatori, A. Mandarino, and M. G. A. Paris, Phys. Rev. A \textbf{40}, 022111 (2014).

\bibitem{Felicetti2020} L. Garbe, M. Bina, A. Keller, M. G. A. Paris, and S. Felicetti, Phys. Rev. Lett. \textbf{124}, 120504 (2020).

\bibitem{Chu2021} Y. Chu, S. Zhang, B. Yu, and J. Cai, Phys. Rev. Lett. \textbf{126}, 010502 (2021).

\bibitem{Rams} M. M. Rams, P. Sierant, O. Dutta, P. Horodecki, and J. Zakrewski, Phys. Rev. X \textbf{8}, 021022 (2022).

\bibitem{Pezze} L. Pezz\'e, A. Trenkwalder, and M. Fattori, arXiv:1906.01447.

\bibitem{Ying2022} Z.-J. Ying, S. Felicetti, G. Liu, and D. Braak, Entropy \textbf{24}, 1015 (2022).

\bibitem{Gietka2021} K. Gietka, F. Metz, T. Keller, and J. Li, Quantum \textbf{5}, 489 (2021).

\bibitem{Ding} D.-S. Ding, Z.-K. Liu, B.-S. Shi, G.-C. Guo, K. Molmer, and C. S. Adams, Nat. Phys. \textbf{18}, 1447 (2022).

\bibitem{Zhou2023} L. Zhou, J. Kong, Z. Lan, and W. Zhang, Phys. Rev. Research \textbf{5}, 013087 (2023).

\bibitem{Kessler2012} E. M. Kessler, G. Giedke, A. Imamoglu, S. F. Yelin, M. D. Lukin, and J. I. Cirac, Phys. Rev. A \textbf{86}, 012116 (2012).

\bibitem{Minganti2018} F. Minganti, A. Biella, N. Bartolo, and C. Ciuti, Phys. Rev. A \textbf{98}, 042118 (2018).

\bibitem{Fitzpatrick2017} M. Fitzpatrick, N. Sundaresan, A. C. Y. Li, J. Koch, and A. A. Houck, Phys. Rev. X \textbf{7}, 011016 (2017).

\bibitem{Fink2017} T. Fink, A. Schade, S. H\"ofling, C. Schneider, and A. Imamoglu, Nat. Phys. \textbf{14}, 365 (2018).

\bibitem{Li2022} Z. Li, F. Claude, T. Boulier, E. Giacobino, Q. Glorieux, A. Bramati, and C. Ciuti, Phys. Rev. Lett. \textbf{128}, 093601 (2022).

\bibitem{Benary2022} J. Benary, C. Baals, E. Bernhart, J. Jiang, M. R\"ohrle, and H. Ott, New J. Phys. \textbf{24}, 103034 (2022).

\bibitem{Lorenzo2017} S. Fern\'andez-Lorenzo and D. Porras, Phys. Rev. A \textbf{96}, 013817 (2017).

\bibitem{Lorenzo2018} S. Fern\'andez-Lorenzo, J. A. Dunningham, and D. Porras, Phys. Rev. A \textbf{97}, 023843 (2018).

\bibitem{Ivanov2020} P. A. Ivanov, Phys. Scr. \textbf{95}, 025103 (2020).

\bibitem{Ivanov2020_1} P. A. Ivanov, Phys. Rev. A \textbf{102}, 052611 (2020).

\bibitem{Ilias2022} T. Ilias, D. Yang, S. F. Huelga, and M. B. Plenio, PRX Quantum, \textbf{3}, 010354 (2022).

\bibitem{Macieszczak2016} K. Macieszczak, M. Guta, I. Lesanovsky, and J. P. Garrahan, Phys. Rev. A \textbf{93}, 022103 (2016).

\bibitem{Munoz} C. S. Munoz, B. Buca, J. Tindall, A. Gonz\'alez-Tudela, D. Jaksch, and D. Porras, arXiv:1903.05080.

\bibitem{Paris2009} M. G. A. Paris, Int. J. Quant. Inf. \textbf{7}, 125 (2009).

\bibitem{Breuer} H. P. Breuer and F. Petruccione, \emph{The Theory of Open Quantum Systems} (Oxford University Press, Oxford, 2007).

\bibitem{Torre2013} E. G. D. Dalla Torre, J. Otterbach, E. Demler, V. Vuletic, and M. D. Lukin, Phys. Rev. Lett. \textbf{110}, 120402 (2013).

\bibitem{Tudela2013} A. Gonz\'ales-Tudela and D. Porras, Phys. Rev. Lett. \textbf{110}, 080502 (2013).

\bibitem{Groszkowski2022} P. Groszkowski, M. Koppenh\"ofer, H.-K. Lau, and A. A. Clerk, Phys. Rev. X \textbf{12}, 011015 (2022).

\bibitem{Puri1979} R. Puri and S. Lawande, Phys. Lett. A \textbf{72}, 200 (1979).

\bibitem{Lawande1981} S. Lawande, R. Puri, and S. Hassan, J. Phys. B: At. Mol. Phys. \textbf{14}, 4171 (1981).

\bibitem{Pezze2009} L. Pezz\'e and A. Smerzi, Phys. Rev. Lett. \textbf{102}, 100401 (2009).

\bibitem{Weedbrook2012} C. Weedbrook \emph{et al}., Rev. Mod. Phys. \textbf{84}, 621 (2012).

\bibitem{Pinel2013} O. Pinel, P. Jian, N. Treps, C. Fabre, and D. Braun, Phys. Rev. A \textbf{88}, 040102(R) (2013).

\bibitem{Wineland} D. J. Wineland, J. J. Bollinger, W. M. Itano, F. L. Moore, and D. J. Heinzen, Phys. Rev. A \textbf{50}, 67 (1994).

\bibitem{Ma2011} J. Ma, X. Wang, C. P. Sun, and F. Nori, Phys. Rep. \textbf{509}, 89 (2011).

\bibitem{LMG_model} J. Ma and X. Wang, Phys. Rev. A \textbf{80}, 012318 (2009).






\end{thebibliography}
\end{document}